\documentclass[twocolumn,prb,showpacs,preprintnumbers]{revtex4}
\usepackage{graphicx}
\usepackage{dcolumn}
\usepackage{bm}
\usepackage{float}

\begin{document}

\title{Dark Periods in Rabi Oscillations of Superconducting Phase Qubit Coupled to a Microscopic
Two-Level System}

\author{ Xueda Wen$^1$, Shi-Liang Zhu$^2$, and Yang Yu$^1$}\email{ yuyang@nju.edu.cn}

\affiliation{$^1$National Laboratory of Solid State
Microstructures and Department of Physics, Nanjing University,
Nanjing 210093, China \\
$^2$Laboratory of Quantum Information Technology, ICMP and SPTE,
South China Normal University, Guangzhou, China}

\begin{abstract}

We propose a scheme to demonstrate macroscopic quantum jumps in a
superconducting phase qubit coupled to a microscopic two-level
system in the Josephson tunnel junction. Irradiated with suitable
microwaves, the Rabi oscillations of the qubit exhibit signatures
of quantum jumps: a random telegraph signal with long intervals of
intense macroscopic quantum tunneling events (bright periods)
interrupted by the complete absence of tunneling events (dark
periods). An analytical model is developed to describe the width
of the dark periods quantitatively. The numerical simulations
indicate that our analytical model can capture underlying physics
of the system. Besides calibrating the quality of the microscopic
two-level system, our results have significance in quantum
information process since dark periods in Rabi oscillations are
also responsible for errors in quantum computing with
superconducting  qubits.

\end{abstract}

\maketitle

\section{INTRODUCTION}

Superconducting Josephson devices coherently driven by external
fields provide new insights into fundamentals of quantum mechanics
and hold promise for use in quantum computation and quantum
information.\cite{Mooij,Makhlin,Nielsen} Recent experiments based
on Josephson tunnel junctions have unambiguously demonstrated the
quantum behavior of macroscopic variable.
\cite{Jackel,Nakamura,Yu,Martinis} One interesting quantum
phenomenon is known as quantum jumps, which was proposed by Bohr
as early as 1913.\cite{Bohr} Bohr suggested that the interaction
of light and matter occurs in such a way that an atom undergoes
instantaneous transitions of its internal state upon the emission
or absorption of a light quantum. These sudden transitions have
become known as 'quantum
jumps'.\cite{Scully,Orszag,Carmichael,Plenio,Blatt} Quantum jumps
were firstly observed in experiments in the 1980s by investigating
the fluorescence of a single trapped ion driven by
laser.\cite{ion} In such atomic systems, quantum jumps are a
random telegraphic process with long intervals of intense photon
emissions interrupted by periods of the absence of photons. To
observe quantum jumps in macroscopic systems such as
superconducting qubits, one may naturally think that it can be
achieved by detecting microwave photon emissions in analogy with
the way in atomic systems. However, this idea suffers from the
absence of microwave photodetectors, although some theoretical
efforts have been made.\cite{Helmer,Romero} Recently, macroscopic
quantum jumps were experimentally demonstrated for the first time
in a superconducting phase qubit coupled with a TLS inside the
Josephson junction.\cite{Yu2} In this experiment, the state of the
system is read out not by detecting photon emissions but by
detecting macroscopic quantum tunneling events, and quantum jumps
behave in the form of jumping randomly between upper branch and
lower branch of the switching currents. In addition, recent
experiments on superconducting charge qubits also demonstrated
quantum jumps by observing quasiparticle tunneling in the time
domain,\cite{Aumentado,Shaw} and theory for the kinetics of the
system has been developed.\cite{Lut}

In this article, we show that macroscopic quantum jumps can be
better observed in a Rabi-oscillation experiment in the phase
qubit-TLS coupling system.
The main points we are going to make are the following. Firstly,
quantum jumps in coherent excitations of macroscopic quantum
states can be well studied based on our model. In the experiment
by Yu \textit{et al}.,\cite{Yu2} the energy level structure keeps
changing during the measurement time, and coherent characteristic
of the dynamics cannot be observed directly. The scheme proposed
in this article is based on a fixed energy level structure, which
is usually implemented to demonstrate Rabi oscillations as the
basic skill to manipulate quantum states. As we can see below, as
a result of the coherent dynamics in our model, some new features
of quantum jumps which have not been expected in the previous
experiments can appear.

Secondly, macroscopic quantum jumps in the qubit-TLS coupling
system bring new insights into our understanding of the effects of
TLS on the dynamics of the qubit. Qubit-TLS coupling system has
received dramatic attention recently in both
experimental\cite{Simmonds,Kim,Lup} and
theoretical\cite{Shnirman,Faoro,Clare1,Galperin,Clare2,Ashhab}
studies because it is suggested that TLS is a major source of
decoherence in superconducting Josephson qubits. In particular,
Rabi oscillations in such qubit-TLS coupling system have been
studied by several authors. \cite{Simmonds,Galperin,Clare2,Ashhab}
However, the previous works mainly emphasized on the ensemble
characteristic of the system, without considering the effects of
quantum jumps in single trajectories, which is actually the
original form of experimental data.\cite{Yu,Yu2} Therefore, a
detailed description of the effect of TLS on the trajectories of a
single qubit is desirable.

Thirdly, recent experiments have demonstrated that TLS can be used
as quantum memory with good performance.\cite{Neeley} It is also
suggested that such TLSs themselves can serve as qubits
\cite{Zagoskin,LinTian} and can be implemented to generate genuine
multi-qubit entangled states.\cite{Zhu} Therefore, characterizing
the TLS is critical to improve the performance of such solid-state
qubits. In this article, we show that the macroscopic quantum
jumps phenomenon can be used to read out the state of TLS, which
provides a new approach to calibrate individual TLS
quantitatively.

This article is organized as follows. In Sec. II we briefly
describe the basic physics of superconducting phase qubit under
coherent driving fields, and then introduce the Monte Carlo
wavefunction method which can be adopted to study the Rabi
oscillations in the superconducting phase qubit. In Sec. III we
introduce the qubit-TLS coupling system, and present the
Hamiltonian of the hybrid system. In Sec. IV, we provide a clear
physical picture of macroscopic quantum jumps in such qubit-TLS
coupling system. In Sec. V, we study the width of dark periods in
Rabi oscillations quantitatively with both numerical and
analytical methods. In Sec VI, we show that the quantum jumps
approach can be used to characterize the TLS quantitatively, and
this article ends with a brief discussion on the relation between
quantum jumps and readout fidelity in superconducting quantum
computing.

\section{QUANTUM JUMP APPROACH TO RABI OSCILLATIONS IN SUPERCONDUCTING PHASE QUBIT}

In this section we briefly present the basic physics of
superconducting Josephson phase qubit driven by coherent fields,
and then introduce how the quantum-jump approach can be used to
simulate Rabi oscillations of a single qubit. To grasp the spirit
of quantum-jump approach in a simple way, here we firstly consider
the qubit without coupling to a TLS.

Superconducting Josephson phase qubit is essentially a current-biased Josephson junction.
The Hamiltonian of the phase qubit as shown in Fig.1(a) reads\cite{Leggett,Martinis2,Clarke,Martinis3}
\begin{equation}\label{Qubit Hamiltonion}
H_{qb}=\frac{1}{2C}\hat{Q}^{2}-\frac{I_{0}\Phi_{0}}{2\pi}\cos\hat{\delta}
-\frac{I\Phi_{0}}{2\pi}\hat{\delta},
\end{equation}
where $I_{0}$ is the critical current of the Josephson junction,
$I$ is the bias current, $C$ is the junction capacitance,
$\Phi_{0}=h/2e$ is the flux quantum, $\hat{Q}$ denotes the charge
operator and $\hat{\delta}$ represents the gauge invariant phase
difference across the junction, which obeys the convectional
quantum commutation relation $[\hat{\delta},\hat{Q}]=2ei$. Quantum
behavior can be observed for large area junctions when the bias
current is slightly smaller than the critical current. The
junction works as a phase qubit when the Josephson coupling
energy $E_{J}=I_{0}\Phi_{0}/2\pi$ is much larger than the single
charging energy $E_{C}=e^{2}/2C$. In this regime, the two lowest
energy levels, $|0\rangle$ and $|1\rangle$ as shown in Fig.1(b),
are usually employed as a qubit in quantum computation. The state
of the qubit can be controlled through the bias current $I(t)$
given by\cite{Martinis3}
\begin{equation}\label{control}
I(t)=I_{dc}+\Delta I(t)=I_{dc}+ I_{\mu w}\cos\omega t,
\end{equation}
where the classical bias current is parameterized by the dc
current $I_{dc}$ and the ac current with the magnitude $I_{\mu w}$
and frequency $\omega$.

Truncating the full Hilbert space of the junction to the qubit
subspace $\{ |0\rangle$, $|1\rangle\}$, the Hamiltonian of the
phase qubit can be written as
\begin{equation}\label{qubit Hamiltonian}
H_{qb}=\hbar\omega_0|0\rangle\langle0| +\hbar\omega_1| 1\rangle\langle 1|
+\hbar\Omega_m\cos\omega t(|0\rangle\langle1|+|1\rangle\langle0|),
\end{equation}
where $\omega_n$ is the energy frequency of state $|n\rangle$,
$\Omega_m =I_{\mu w}\sqrt{1/2\hbar\omega_{10}C}$ is Rabi
frequency, and $\omega_{10}=\omega_1-\omega_0$ is the energy
frequency between state $|0\rangle$ and $|1\rangle$. In the
interaction picture and choosing a rotating frame of the frequency
$\omega$, Hamiltonian (\ref{qubit Hamiltonian}) can be simplified
to
\begin{equation}\label{qubit Hamiltonian 2}
H_{qb}=\hbar\Delta|1\rangle\langle1|
+\frac{\hbar\Omega_m}{2}(|0\rangle\langle1|+|1\rangle\langle0|),
\end{equation}
where $\Delta\equiv\omega_{10}-\omega$ represents the detuning.

\begin{figure}
\centering
\includegraphics[width=3.0375in]{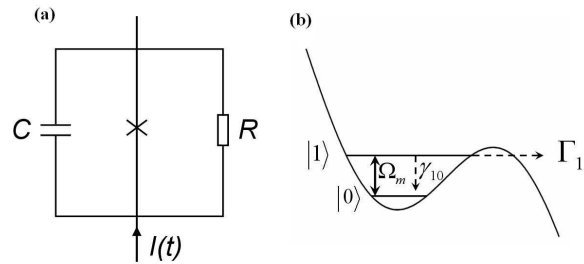}
\caption{(a) Equivalent circuit of a current-biased Josephson tunnel
junction. (b) An illustration of various coherent and incoherent processes in a superconducting
phase qubit radiated by a microwave.}
\end{figure}

To better understand our simulation method, we give a brief description of the experimental
procedure firstly. In experiments,\cite{Yu} the system is prepared in the initial state
$|0\rangle$ firstly, then a microwave source is turned on for a duration time $\tau_m$
(For convenience, we call $\tau_m$ the measurement time below).
Because the tunneling rates depend exponentially on the barrier height, the bias current
$I_{dc}$ can be chosen so that the tunneling from $|0\rangle$ is essentially 'frozen out'.\cite{Yu}
Therefore, the Rabi oscillations between states $|0\rangle$ and $|1\rangle$ lead to an
oscillating probability for the system to tunnel out of the potential. At the end of
the measurement time $\tau_m$ for a single run, no matter the tunneling event has happened
or not, the biased current $I_{dc}$ is adjusted to initialize the state for the
next run. Here we label the time interval for initializing the qubit state as $\tau_p$.

To simulate the dynamics of the phase qubit system with and
without TLS coupling, we adopt Monte Carlo wavefunction method,
which is also called 'quantum trajectories' method\cite{Scully,
Orszag,Carmichael,Plenio,Dalibard,Dalibard2,Carmichael2}. In this
method, the time evolution of the system can be described by the
non-Hermitian effective Hamiltonian
\begin{equation}\label{effective Hamiltonian}
H_{eff}=H_{qb}-\frac{i\hbar}{2}(\gamma_{10}+\Gamma_1)|1\rangle\langle1|,
\end{equation}
where $\gamma_{10}$ is the rate of energy relaxation from
$|1\rangle$ to $|0\rangle$ and $\Gamma_1$ is the tunneling rate
from the level $|1\rangle$ out of the potential. The 'conditional
state' $|\Psi_c(t)\rangle$ of the system is determined by
\begin{equation}\label{conditional state vector}
|\Psi_c(t+\delta t)\rangle=\exp[-iH_{eff}\delta t/\hbar]|\Psi_c(t)\rangle.
\end{equation}
The procedure adopted in quantum-jump simulation of Rabi
oscillations of a single qubit can be summarized as follows:

\begin{figure}
\centering
\includegraphics[width=3.3in]{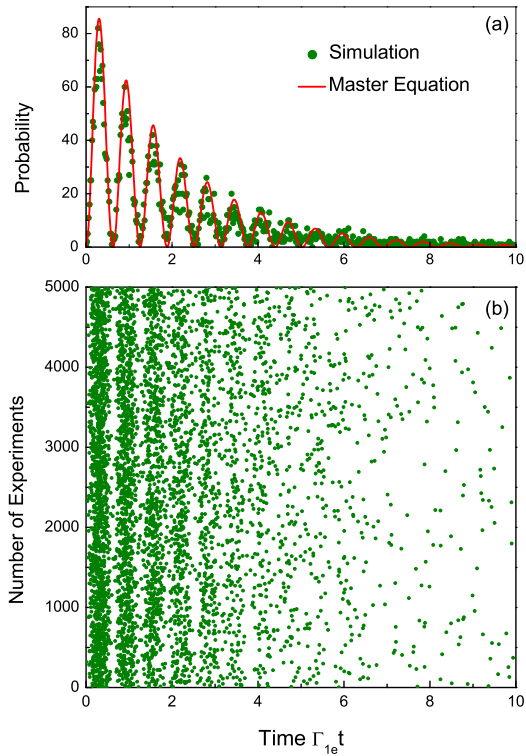}
\caption{(Color online)Simulated Rabi oscillations of a single
superconducting phase qubit. (a) Average over $N=5000$ quantum
tunneling events obtained using Monte Carlo wavefunction
method(green dots). The red line is an ensemble result obtained
using master equation. (b) Trajectories of quantum tunneling
events obtained with Monte Carlo wavefunction method. The
parameters are $\Omega_m=10\Gamma_1$, $\Delta=0$, and
$\gamma_{10}=\Gamma_1/4$. }
\end{figure}

(i) Determine the current probability of a relaxation or tunneling event, i.e.,
$\Delta P_r=\gamma_{10}\delta t \langle \Psi|1\rangle\langle1|\Psi\rangle$ and
$\Delta P_t=\Gamma_1\delta t \langle \Psi|1\rangle\langle1|\Psi\rangle$.

(ii) Obtain random numbers $r_1$ and $r_2$ between zero and one, compare with $\Delta P_t$
and $\Delta P_r$ respectively, and decide on tunneling or relaxation for different cases:

(a) If $r_1<\Delta P_t$ a quantum tunneling event happens. Register the time $t$ of the event,
and then turn to step (iii).

(b) If $r_1>\Delta P_t$ and $r_2<\Delta P_t$ there is a relaxation, so that the system
jumps to the state $|0\rangle$.

(c) If $r_1>\Delta P_r$ and $r_2>\Delta P_t$ no quantum jumps take place, so the system evolves
under the influence of the non-Hermitian form
\begin{equation}
|\Psi\rangle\to \frac{\{1-(i/\hbar)H_{qb}\delta t-(\gamma /2)\delta t|1\rangle\langle 1|\}|\Psi\rangle}{(1-\Delta P)^{1/2}},
\end{equation}
where $\gamma\equiv\gamma_{10}+\Gamma_1$ and $\Delta P\equiv
\Delta P_r+\Delta P_t$.

(iii) If no tunneling event happens, repeat the previous steps
until the end of the measurement time $\tau_m$.


(iv) Accounting time $t$ over many simulation runs.

By using this Monte Carlo wavefunction method,  we simulate the trajectories of Rabi
oscillations in a single superconducting phase qubit and the
results are shown in Fig.2(b). We emphasize that it is possible
that a tunneling event never happens during a finite measurement
time $\tau_m$ for a single run, which is important for observing
dark periods as discussed in Sec. V. To test the efficiency of our
simulation method, we compare the simulated results with those
obtained with master equation used in Ref [6]. As shown in
Fig.2(a), they agreed very well. Actually, it can be proved that
the Monte Carlo method is equivalent, on average, to the master
equation.\cite{Scully,Orszag,Carmichael} Guaranteed with the
efficient method for simulating Rabi oscillations in
superconducting phase qubit, we turn to the more interesting
qubit-TLS coupling system.

\section{HAMILTONIAN OF QUBIT-TLS COUPLING SYSTEM}

Experiments have shown that some TLSs may locate inside the Josephson tunnel barrier,
as illustrated  in Fig.3(a). A TLS is understood to be an atom, or a small group of atoms,
that tunnels between two lattice configurations, \cite{Phillips,Neeley}, with different
wave functions $|L\rangle$ and $|R\rangle$. The two states of the TLS correspond to two
different values of the Josephson junction critical current $I_0$ which is proportional to
the square of the tunneling matrix element across the junction. When the TLS is in state
$|R\rangle$ ($|L\rangle$), the junction critical current is $I_{0R}$ ($I_{0L}$).
Then the interaction Hamiltonian between the qubit and the TLS is:
\begin{equation}\label{interaction Hamiltonian}
H_{int}=-\frac{\Phi_0 I_{0R}}{2\pi}\cos\delta\otimes|R\rangle\langle R|
-\frac{\Phi_0 I_{0L}}{2\pi}\cos\delta\otimes|L\rangle\langle L|.
\end{equation}

\begin{figure}
\centering
\includegraphics[width=3.3375in]{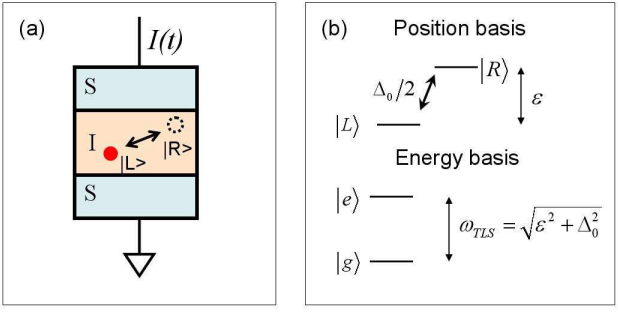}
\caption{(Color online) (a) Schematic of a TLS in the oxide tunnel barrier of the Josephson
junction. Because the barrier material is disordered, some atoms can occupy two positions, labelled
$|L\rangle$ and $|R\rangle$. (b) The position states are separated by an energy difference $\varepsilon$
and connected by a tunneling energy $\Delta_0/2$. In the energy eigenstate basis, the ground state
$|g\rangle$ and excited state $|e\rangle$, are separated by an energy $\omega_{TLS}$. The charge motion
between $|L\rangle$ and $|R\rangle$ couples to the currents and voltages of the qubit circuit.}
\end{figure}

Assume an asymmetric potential with energy separated by
$\hbar\omega_{TLS}$ for the TLS, then the ground and excited
states are
$|g\rangle=\sin(\theta/2)|L\rangle+\cos(\theta/2)|R\rangle$ and
$|e\rangle=\cos(\theta/2)|L\rangle-\sin(\theta/2)|R\rangle$, where
$\omega_{TLS}=\sqrt{\varepsilon^2+\Delta_0^2}$,
$\theta=\arctan(\Delta_0/\varepsilon)$ with $\varepsilon/2$ being
the energy asymmetry and $\Delta_0/2$ being the bare tunneling
matrix element (Fig.3(b)). Considering the junction is biased near
its critical current\cite{Martinis3,Clare2}, i.e., $\delta
=\pi/2-\delta'$ with $|\delta'| \ll 1$, then
$\cos(\pi/2-\delta')=\sin\delta'\approx\delta'$. In this case,
$\delta'$ can be well approximated as position coordinate operator
of the harmonic oscillator. Using these facts and including only
the dominant resonant terms arising from the interaction
Hamiltonian, Equation (\ref{interaction Hamiltonian}) becomes
\begin{equation}\label{interaction Hamiltonian2}
H_{int}=\frac{\delta I_0\sin\theta}{2}\sqrt{\frac{\hbar}{2\omega_{10}C}}(|0e\rangle\langle1g|+|1g\rangle\langle0e|),
\end{equation}
where $\delta I_0\equiv I_{0R}-I_{0L}$ is the fluctuation
amplitude in $I_0$ produced by the TLS. The coupling of the two
intermediate energy levels through $H_{int}$ produces an energy
splitting which can be characterized in spectroscopic
measurements.\cite{Yu2,Simmonds,Kim,Lup} The Hamiltonian of the
qubit-TLS coupling system in the basis \{$|0g\rangle$,
$|1g\rangle$, $|0e\rangle$, and $|1e\rangle$\} then reads
\begin{equation} \label{origin Hamiltonian}
H=\hbar
\left(\begin{array}{cccc}
 0                      & \Omega_m \cos\omega t    &0             &0\\
\Omega_m \cos\omega t   & \omega_{10}              &\Omega_c      &0\\
0                      & \Omega_c                 & \omega_{TLS} & \Omega_m \cos\omega t\\
0                      & 0                        & \Omega_m \cos\omega t   & \omega_{10}+\omega_{TLS}\\
\end{array} \right).
\end{equation}

To understand the underlying physics that governs the dynamic of
the system more clearly, we chose the interaction picture and make
a transformation to a rotating frame. Then the Hamiltonian can be
simplified to the time independent form (see appendix)
\begin{equation}\label{simple form}
H'=\hbar
\left(\begin{array}{cccc}
 0                      & \Omega_m/2                &0                        &0\\
\Omega_m/2              & \Delta                    &\Omega_c                 &0\\
0                       & \Omega_c                  & \Delta +\Delta_r        & \Omega_m/2\\
0                       & 0                         & \Omega_m/2              & 2\Delta +\Delta_r \\
\end{array} \right),
\end{equation}
where $\Delta\equiv\omega_{10}-\omega$ and
$\Delta_r\equiv\omega_{TLS}-\omega_{10}$ are the detunings,
$\Omega_c=(\delta I_0\sin\theta/2)\sqrt{1/(2\hbar\omega_{10}C)}$ i
s the avoided energy level crossing between $|0e\rangle$ and
$|1g\rangle$, and $\Omega_m =I_{\mu w}\sqrt{1/2\hbar\omega_{10}C}$
is the Rabi frequency between the qubit state $|0\rangle$ and
$|1\rangle$. From Hamiltonian (\ref{simple form}), we can divide
the system into two subspaces $A\equiv\{|0g\rangle, |1g\rangle\}$
and $B\equiv\{|0e\rangle, |1e\rangle\}$. The effective coupling
strength between $A$ and $B$ is decided by several parameters
including $\Omega_c$, $\Delta_r$, $\Delta$ and $\Omega_m$.
Interestingly, the parameters $\Delta_r$, $\Delta$ and $\Omega_m$
can be easily controlled by adjusting the bias current $I_{dc}$,
the microwave frequency and the microwave amplitude,
respectively.\cite{Yu2} While in the interior of the subsystem,
the coupling strength $\Omega_m/2$ between states $|0g\rangle$ and
$|1g\rangle$ ($|0e\rangle$ and $|1e\rangle$ ) can be controlled by
adjusting the amplitude of the microwave.\cite{Yu, Martinis}
Therefore, the qubit-TLS coupling system behaves as an "artificial
atom" which can be manipulated flexibly.

\section{QUANTUM JUMPS IN QUBIT-TLS COUPLING SYSTEM}

To have a clear physical picture of macroscopic quantum jumps in
qubit-TLS coupling system, we provide an analogy with quantum
jumps in quantum optics, which has been well developed for years.
\cite{Scully,
Orszag,Carmichael,Plenio,Blatt,ion,Dalibard,Dalibard2,Carmichael2}
In quantum optics, the system generally has energy structure shown
in Fig.4(a). Two excited states $|1\rangle$ and $|2\rangle$ are
connected to a common ground state $|0\rangle$ via a strong and
weak transition, respectively. The fluorescent photons from the
strong transition are observed. However, an excitation of the weak
transition where the electron is temporarily shelved in the
metastable level $|2\rangle$ will cause the strong transition to
be turned off. Therefore, it is possible to monitor the quantum
jumps of the weak transition via the signal provided by the
fluorescence of the strong transition. In the language of quantum
measurement theory,\cite{Zurek} the fluorescence from the strong
transition acts as a pointer from which the microscopic quantum
state of the atom may be determined.

\begin{figure}
\centering
\includegraphics[width=3.3375in]{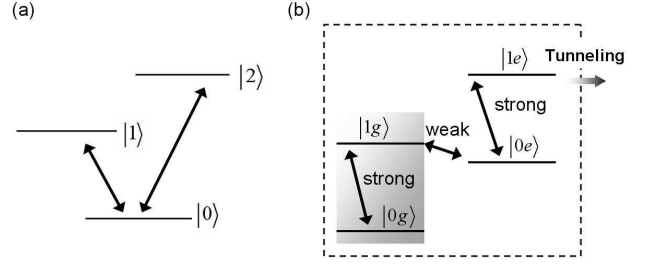}
\caption{(a) $V$-type energy structure for observing quantum jumps in the fluorescence of a single atom.
Two upper levels $|1\rangle$ and $|2\rangle$ couple to a common ground state $|0\rangle$. The $|1\rangle
\leftrightarrow |0\rangle$ transition is assumed to be strong while the $|2\rangle\leftrightarrow|0\rangle$
transition is weak.
(b) Schematic energy level diagram for a junction coupled to a TLS. The ground state (the excited state)
of the qubit is denoted as $|0\rangle$ ($|1\rangle$); $|g\rangle$ and $|e\rangle$ represent the ground state
and the excited state of TLS. The $|1g\rangle \leftrightarrow |0e\rangle$ transition is weak, while the
$|0e\rangle\leftrightarrow|1e\rangle$($|0g\rangle\leftrightarrow|1g\rangle$) transition is strong.
}
\end{figure}

Similarly, in the qubit-TLS coupling system considered here, quantum jumps between macroscopic
quantum states are also proposed to happen. As discussed in Sec. III, we can have a weak
transition between states $|1g\rangle$ and $|0e\rangle$ by adjusting the
bias current and a strong transition between $|0e\rangle$ and $|1e\rangle$(or between
$|0g\rangle$ and $|1g\rangle$) by adjusting the microwave amplitude. However, different
from observing fluorescent photons in atom systems, the states of the qubit-TLS coupling
system are read out through quantum tunneling. As has been proved in experiments \cite{Yu2},
in the qubit-TLS coupling system, different states correspond to different tunneling rates.\cite{reason1}
Therefore, by adjusting the bias current at a proper value, only the tunneling from state
$|1e\rangle$ is prominent, and tunneling from other states can be neglected.

\begin{figure}
\centering
\includegraphics[width=3.3375in]{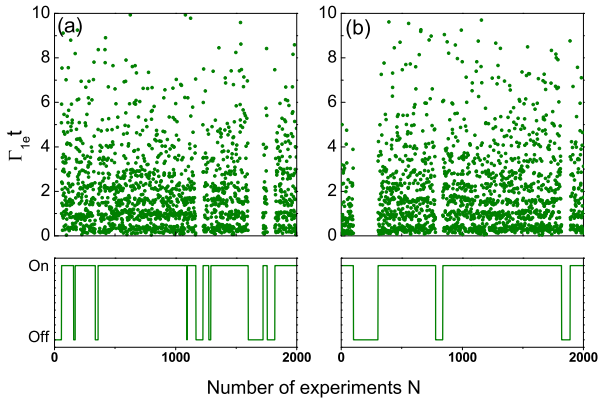}
\caption{(Color online) Dark periods in Rabi oscillations of qubit-TLS coupling system.
 The detunings are (a) $\Delta_r=2\Gamma_{1e}$
and (b) $\Delta_r=5\Gamma_{1e}$. Lower panel: the filtered data
corresponds to on and off of the Rabi oscillations. The parameters
for simulation are $\Delta=0$, $\gamma_{10}=\Gamma_{1e}/4$,
$\Omega_c=\Gamma_{1e}/4$, $\Omega_m=10\Gamma_{1e}$, and
$\tau_m=10/\Gamma_{1e}$.} \label{FIG.5}
\end{figure}

Then the physics for quantum jumps in qubit-TLS coupling system is
straightforward. When the system stays in subspace
$B\{|0e\rangle,|1e\rangle\}$, quantum tunneling events in the Rabi
oscillation form (see Fig.2) can be observed. However, once the
system transitions into subspace $A\{|0g\rangle,|1g\rangle\}$
(dark area in Fig.4(b)), where the system has little probability
to tunnel out the potential, no quantum tunneling events can be
observed during the measurement time, and then dark periods in
Rabi oscillations appear. To substantiate our proposal, we
investigate this phenomena numerically with the Monte Carlo
wavefunction method introduced in Sec. II. In the qubit-TLS
coupling system, the dynamics can be described by the
non-Hermitian effective Hamiltonian
\begin{equation}\label{Effective fourlevel Hamiltonian}
H_{eff}=H'-\frac{i\hbar}{2}(\gamma_{10}+\Gamma_{1e})|1e\rangle\langle1e|-\frac{i\hbar}{2}\gamma_{10}|1g\rangle\langle1g|,
\end{equation}
where $\Gamma_{1e}$ is the tunneling rate of state $|1e\rangle$
and $\gamma_{10}$ is the relaxation rate from $|1e\rangle$ to
$|0e\rangle$ ($|1g\rangle$ to $|0g\rangle$). It is noticed that we
do not take into account the relaxation effect of the TLS, because
the lifetime of the TLS is much longer than that of the phase
qubit.\cite{Zhu,Zagoskin,Simmonds} Actually, this effect can be
easily considered in the same way as we consider the relaxation of
the qubit. The quantum state either evolves according to the
schr\"odinger equation, or 'jumps' to an eigenstate of the system
with certain probability. The measurement result serves as a
pointer to determine the quantum state of the system. Once a
quantum tunneling event happens, we know that the state of TLS is
$|e\rangle$; then in the next run the initial state of the system
is prepared in the corresponding qubit ground state $|0e\rangle$.
On the contrary, if no quantum tunneling event happens during a
sufficient long measurement time $\tau_m$, we can confirm that the
system is in subspace $A\{|0g\rangle,|1g\rangle\}$, i.e., the
state of TLS is $|g\rangle$. Then in the next run of simulation
the initial state is prepared in the corresponding qubit ground
state $|0g\rangle$ and dark periods in Rabi oscillations may
appear. We emphasize that if the measurement time $\tau_m$ is
short, the TLS's state cannot be determined if no quantum
tunneling event happens at the end of $\tau_m$. In this case, we
decide the initial state of the next run using Monte Carlo method
according to the population of $|g\rangle$ and $|e\rangle$
respectively.

As shown in Fig.5, in the simulated trajectories the quantum
tunneling events are disturbed by dark periods during which no
tunneling events are observed. In addition, it is found that the
average width $\langle N_D\rangle$ of dark periods for
$\Delta_r=2\Gamma_{1e}$ is smaller than that for
$\Delta_r=5\Gamma_{1e}$ (Here $\langle N_D\rangle$ means the the
average number of runs in a single dark period). One may simply
think that as the detuning $\Delta_r$ increases, the coupling
between subspace $A$ and subspace $B$ becomes weaker. Therefore
the system is more difficult to make transitions between $A$ and
$B$, thus leading to a larger average width of dark periods.
However, this is not always the case. As discussed below, we will
give a quantitative description between $\langle N_D\rangle$ and
$\Delta_r$.

\section{RELATION BETWEEN DARK PERIODS AND DETUNING }

 \begin{figure}
\centering
\includegraphics[width=3.3375in]{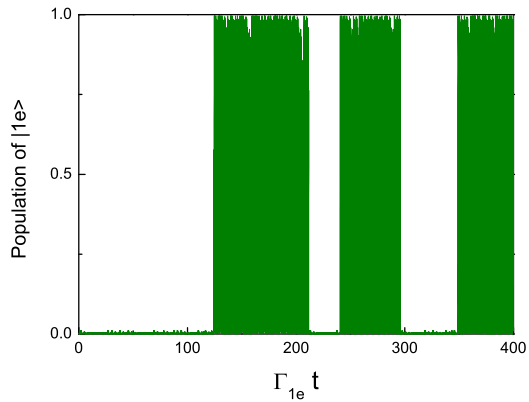}
\caption{(Color online) The time evolution of the population of
state $|1e\rangle$. The dark periods correspond to the shelving of
the system in subspace $A\{|0g\rangle, |1g\rangle\}$. The
parameters are $\Delta=0$, $\gamma_{10}=\Gamma_{1e}/4$,
$\Omega_c=\Gamma_{1e}/4$, $\Omega_m=10\Gamma_{1e}$, and
$\Delta_r=2\Gamma_{1e}$ }
\end{figure}

In this section we study the relationship between the average
width $\langle N_D\rangle$ of dark periods and the detuning
$\Delta_r$. We start from investigating the transition rate from
subspace $A\{|0g\rangle, |1g\rangle\}$ to subspace $B\{|0e\rangle,
|1e\rangle\}$, i.e., $\Gamma_D=1/T_D$, where $T_D$ is the lifetime
of the dark periods. A simple method is to study the time
evolution of the population of $|1e\rangle$ in the qubit-TLS
coupling system based on the effective Hamiltonian (\ref{Effective
fourlevel Hamiltonian}). As shown in Fig.6 (Here we have deducted
the initial-state preparing time $\tau_p$), the population of
$|1e\rangle$ is interrupted by random periods of 'darkness'.
Quantum jumps from dark periods to bright periods indicate that
the system jumps from subspace $A$ to subspace $B$. We make an
ensemble average over the lifetime of the dark periods. As shown
in Fig.7, it is interesting that the transition rates reveal three
peaks as a function of the detuning $\Delta_r$. In addition, we
find that the positions for the two side peaks are $\Omega_m$ and
$-\Omega_m$, respectively.

\begin{figure}
\centering
\includegraphics[width=3.7in]{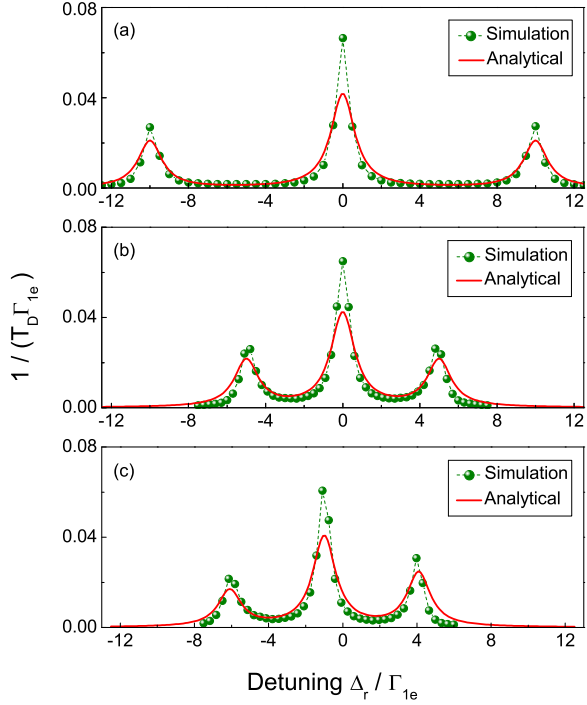}
\caption{(Color online) Transition rate $1/T_D$ versus the
detuning $\Delta_r$. $T_D$ represents the average lifetime of the
dark periods as shown in Fig. 6. The red solid lines show the
analytical results obtained from Eqs.(\ref{W-L}-\ref{detuning})
and the green dots show the simulation results. Each green dot is
obtained by averaging over $N=4000$ dark periods. The parameters
used in both analytical solutions and simulations are:
$\gamma_{10}=\Gamma_{1e}/4$, $\Omega_c=\Gamma_{1e}/4$, and (a)
$\Delta=0$, $\Omega_m=10\Gamma_{1e}$, (b) $\Delta=0$,
$\Omega_m=5\Gamma_{1e}$, (c) $\Delta=1\Gamma_{1e}$,
$\Omega_m=5\Gamma_{1e}$. The positions of the three peaks are
determined by $\Delta_r=-\sqrt{\Delta^2+\Omega_m^2}-\Delta$,
$-\Delta$ and $\sqrt{\Delta^2+\Omega_m^2}-\Delta$, respectively.
It is noticed that the heights of the side peaks are equal for
$\Delta=0$ and unequal for $\Delta\neq 0$. }
\end{figure}

To understand the underlying physics, we rewrite the Hamiltonian
(\ref{simple form}) in the subspace $A$ and $B$, respectively. In
the resonant case $\Delta=0$, the Hamiltonian in the basis
$\{|A_e\rangle\equiv(|0g\rangle+|1g\rangle)/\sqrt{2},
|A_g\rangle\equiv(|0g\rangle-|1g\rangle)/\sqrt{2},
|B_e\rangle\equiv(|0e\rangle+|1e\rangle)/\sqrt{2},
|B_g\rangle\equiv(|0e\rangle-|1e\rangle)/\sqrt{2}\}$ reads (see
appendix)
\begin{equation}\label{A-B form}
H_{A-B}=\hbar
\left(\begin{array}{cccc}
 \Omega_m/2                      & 0                &\Omega_c/2                  &\Omega_c/2 \\
 0                      & -\Omega_m/2               &-\Omega_c/2                 &-\Omega_c/2 \\
\Omega_c/2             &-\Omega_c/2                & \Delta_r+\Omega_m/2        & 0         \\
\Omega_c/2             &-\Omega_c/2                & 0                          & \Delta_r-\Omega_m/2 \\
\end{array} \right),
\end{equation}
from which it is clear that the eigenstates in subspace $A$ can
couple to the eigenstates in subspace $B$ independently with the
coupling strength $\Omega_c/2$. Then the physical picture is clear
as follows. The energy separation between the two eigenstates in
each subspace is equal to the Rabi frequency $\Omega_m$. For the
detuning $\Delta_r=0$, $|A_e\rangle$ is resonant with
$|B_e\rangle$, and $|A_g\rangle$ is resonant with $|B_g\rangle$;
for $\Delta_r=\Omega_m$, only $|A_e\rangle$ is resonant with
$|B_g\rangle$; and for $\Delta_r=-\Omega_m$, only $|A_g\rangle$ is
resonant with $|B_e\rangle$. Note that all coupling strengthes are
$\Omega_c/2$, one would intuitively expect the height of the
central peak to be twice that of a side peak, and it confirms from
the results of the simulation. Furthermore,  we utilize the
Wilcox-Lamb method \cite{Lamb}  to obtain an analytical form of
the transition rate $\Gamma_D$. In the  approximation to the first
order, the transition rate from subspace $A$ to subspace $B$ has a
simple form
\begin{equation}\label{W-L}
\Gamma_D=\frac{\Omega_c^2}{2}\sum_{i\in A,j\in B}\frac{\rho_i\gamma}{\Delta_{ij}^2+\gamma^2},
\end{equation}\
where $\gamma=(2\gamma_{10}+\Gamma_{1e})/2$, $\rho_i$ is the
probability of state $|i\rangle$ (here $\rho_{A_e}=
\rho_{A_g}=1/2$), and $\Delta_{ij}$ are the detunings between
states $|i\rangle$ ($|i\rangle\in\{|A_e\rangle, |A_g\rangle\}$)
and $|j\rangle$ (($|j\rangle\in\{|B_e\rangle, |B_g\rangle\}$) in
the forms $\Delta_{Ae,Be}=\Delta_{Ag,Bg}=\Delta_r$,
$\Delta_{Ae,Bg}=\Delta_r-\Omega_m$, and
$\Delta_{Ag,Be}=\Delta_r+\Omega_m$. As shown in Fig.7(a-b), our
approximate analytical results agree with the simulation results
considerably.

Furthermore, Equation (\ref{W-L}) can be straightforwardly
generalized to the off resonant case $\Delta\neq0$ in the form
\begin{equation}\label{detuning}
\Gamma_D=2\Omega_c^2\sum_{i\in A,j\in B}\frac{\rho_i\lambda_{ij}^2\gamma}{\Delta_{ij}^2+\gamma^2},
\end{equation}\
where $\lambda_{ij}$ represent the coupling coefficients between
states $|i\rangle$ and $|j\rangle$. Here
$|i\rangle\in\{|A_e\rangle\equiv\sin\frac{\alpha}{2}|0g\rangle+\cos\frac{\alpha}{2}|1g\rangle,
|A_g\rangle\equiv\cos\frac{\alpha}{2}|0g\rangle-\sin\frac{\alpha}{2}|1g\rangle\}$,
and
$|j\rangle\in\{|B_e\rangle\equiv\sin\frac{\alpha}{2}|0e\rangle+\cos\frac{\alpha}{2}|1e\rangle,
|B_g\rangle\equiv\cos\frac{\alpha}{2}|0e\rangle-\sin\frac{\alpha}{2}|1e\rangle\}$
with $\alpha=\arctan(\Omega_m/\Delta)$. The coupling coefficients
$\lambda_{ij}$ have the forms (see appendix)
$\lambda_{Ae,Be}=\sin\frac{\alpha}{2}\cos\frac{\alpha}{2}$,
$\lambda_{Ag,Bg}=-\sin\frac{\alpha}{2}\cos\frac{\alpha}{2}$,
$\lambda_{Ag,Be}=-\sin^2\frac{\alpha}{2}$, and
$\lambda_{Ae,Bg}=\cos^2\frac{\alpha}{2}$, respectively. The
detunings $\Delta_{ij}$ have the forms
$\Delta_{Ag,Bg}=\Delta_{Ae,Be}=\Delta_r+\Delta$,
$\Delta_{Ag,Be}=\Delta_r+\Delta+\sqrt{\Omega_m^2+\Delta^2}$, and
$\Delta_{Ae,Bg}=\Delta_r+\Delta-\sqrt{\Omega_m^2+\Delta^2}$, then
the positions of the three peaks are given by
$\Delta_r=-\sqrt{\Delta^2+\Omega_m^2}-\Delta$, $-\Delta$ and
$\sqrt{\Delta^2+\Omega_m^2}-\Delta$, respectively. Conclusively,
there are  two main differences between the resonance case
($\Delta=0$) and the off resonance case ( $\Delta\neq0$). Firstly,
the positions of the three peaks for $\Delta=0$ are symmetric with
respect to $\Delta_r=0$ but unsymmetric for $\Delta\neq0$.
Secondly, in the case $\Delta=0$ the heights of the side peaks are
equal, because all the coupling coefficients between $|i\rangle$
and $|j\rangle$ have the same value $1/2$. However, in the case
$\Delta\neq0$, the heights of the side peaks are unequal because
of the different coupling coefficients $\lambda_{ij}$. As
expected, the above analysis is well demonstrated in Fig. 7(b-c),
where the agreements between numerical simulations and analytical
results are clear.

\begin{figure}
\centering
\includegraphics[width=3.7in]{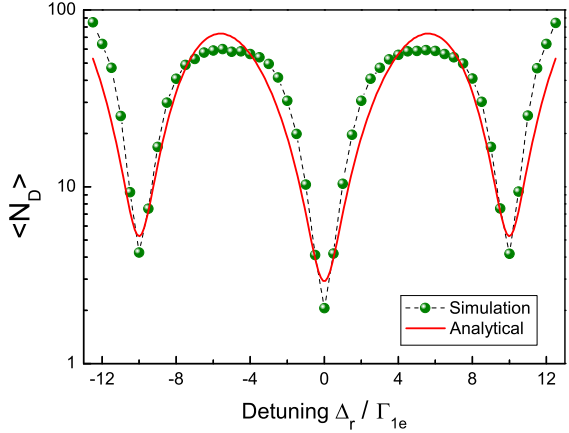}
\caption{(Color online) Average width $\langle N_D\rangle$ of dark periods versus detuning $\Delta_r$.
The red solid line shows the analytical result obtained from Eq.(\ref{W-L})-Eq.(\ref{Average width}),
and the green dot shows the simulation result.
The parameters used in both analytical solution and simulation are: $\Delta=0$, $\gamma_{10}=\Gamma_{1e}/4$, $\Omega_c=\Gamma_{1e}/4$, $\Omega_m=10\Gamma_{1e}$ and $\tau_m=10/\Gamma_{1e}$.
}
\end{figure}

Then we can estimate the average width $\langle N_D\rangle$ of
the dark periods in Fig.5. Supposing the system is in subspace
$A$ initially, the probability for the system residing in $A$
at the end of a single run with interval $\tau_m$ is
\begin{equation}
P=\exp({-\Gamma_D \tau_m}).
\end{equation}
Because the probability $P$ for each run is independent, the
average width of the dark periods can be expressed as
\begin{equation}\label{Average width}
\langle N_D\rangle=\frac{\sum_{n=1}^{\infty}nP^n(1-P)}{\sum_{n=1}^{\infty}P^n(1-P)}=\frac{1}{1-\exp({-\Gamma_D \tau_m})}.
\end{equation}
Both the simulated and analytical results for $\langle N_D\rangle$
are shown in Fig.8 and they agreed very well. The agreement
between the numerical and analytical results indicated the
validity of our model. In experiments, the coupling strength
$\Omega_c$ is typically  20MHz $\sim$
100MHz,\cite{Yu2,Simmonds,Neeley} $\gamma_{10}$ is typically
1MHz$\sim$100MHz, and other parameters including $\Gamma_{1e}$,
$\Delta_r$ and $\Omega_m$ can be controlled flexibly in a large
regime to fulfill our theoretical discussions by manipulating the
biased current or microwave amplitude.\cite{Yu,Martinis}
Therefore, our scheme is feasible within the current technique.

\section{CONCLUSION}

We have proposed a scheme to observe macroscopic quantum jumps in
Rabi oscillations of a superconducting phase qubit coupled to a
TLS.
This scheme provides a new tool to characterize the TLS
quantitatively. Dark periods and bright periods in Rabi
oscillations indicate the TLS residing in states $|g\rangle$ and
$|e\rangle$, respectively. Therefore, the paramerer $\Gamma_D$
discussed in Sec. V describes the transition rate between
$|g\rangle$ and $|e\rangle$ of TLS, which is induced by the
coupling between qubit and TLS. To obtain the intrinsic lifetime
of the TLS, which is known as induced by coupling to elastic
strain field,\cite{Clare2} we just need to decouple the TLS with
qubit by setting the detuning $\Delta_r\gg\Omega_m$. Moreover, in
recent experiments, the TLSs are observed in other kinds of
superconducting qubits such as charge qubit\cite{Aumentado,Shaw,Kim}
and flux qubit\cite{Lup}. We believe that the method and model in
this article can also be generalized to such systems as well as
other systems with similar energy level structure.

Furthermore, in the process of quantum computing, the quantum
computer usually evolves into a highly entangled state. A quantum
jump tends to destroy such entangled state and make the result of
quantum computing incorrect, which is known as decoherence induced
by quantum jumps. However, in this article we found that quantum
jumps can lead to another type of negative effect on quantum
computing, that is, the errors in reading out the quantum state.
For example, once the system jumps into the dark periods, the
qubit's state cannot be read out using the normal methods,
although the qubit can still evolve coherently between state
$|0\rangle$ and $|1\rangle$. In this case, from the measurement
results, we always believe the qubit is staying in state
$|0\rangle$. That is to say, quantum jumps leads to a readout
error in the qubit. Since the previous literatures emphasized the
decoherence induced by the quantum jumps, we hope the discussion
in this article can bring new insights into our understanding of
the effects of quantum jumps on quantum computing.

\section{ ACKNOWLEDGMENTS}

We thank Clare C. Yu for useful discussions. This work was
partially supported by the NSFC (under Contract Nos.
10674062,10725415, and 10674049), the State Key Program for Basic
Research of China (under Contract Nos. 2006CB921801 and
2007CB925204), and the Doctoral Funds of the Ministry of Education
of the People¡¯s Republic of China (under Contract No. 20060284022
).

\section{ APPENDIX: DERIVATION OF EQ.(\ref{simple form}) and EQ.(\ref{A-B form})}

In this appendix, we address the derivation of Eqs. (\ref{simple
form}) and (\ref{A-B form}). Under the interaction picture and the
rotating wave approximation, the Hamiltonian  (\ref{origin
Hamiltonian}) becomes

\begin{equation}\label{rotating approximation}
\hat{H}_I=\hbar
\left(\begin{array}{cccc}
0                             & \Omega_m e^{-i\Delta t}/2                &0                                     &0 \\
 \Omega_m e^{i\Delta t}/2     & 0                                       &\Omega_c e^{-i\Delta_r t}                 &0 \\
0                             &\Omega_c e^{i\Delta_r t}                 & 0        & \Omega_m e^{-i\Delta t}/2         \\
0                             &0               &  \Omega_m e^{i\Delta t}/2                          & 0 \\
\end{array} \right),
\end{equation}
where $\Delta\equiv\omega_{10}-\omega$ and
$\Delta_r\equiv\omega_{TLS}-\omega_{10}$ are the detunings. We
make a transformation to the rotating frame such that the wave
function $|\psi\rangle$ in the original Hamiltonian (\ref{rotating
approximation}) can be expressed as
\begin{equation}\label{transformation}
|\psi\rangle=\hat{U}(t)|\psi'\rangle,
\end{equation}
where
\begin{equation}\label{unitary transformation}
\hat{U}(t)=
\left(\begin{array}{cccc}
1                             & 0                                &0                                     &0 \\
0                             & e^{i\Delta t}                    &0                                     &0 \\
0                             &0                                 & e^{i(\Delta+\Delta_r)t}              &0         \\
0                             &0                                 &0                                     &e^{i(2\Delta+\Delta_r)t} \\
\end{array} \right).
\end{equation}
 Then the Schr$\ddot{o}$dinger equation
\begin{equation}
i\hbar\frac{\partial}{\partial
t}|\psi\rangle=\hat{H}_I|\psi\rangle,
\end{equation}
can now be rewritten as
\begin{equation}
i\hbar\frac{\partial}{\partial
t}|\psi'\rangle=\hat{H}'|\psi'\rangle,
\end{equation}
where
\begin{eqnarray}
\hat{H}'=&&\hat{U}^{\dagger}(t)\hat{H}_I \hat{U}(t)-i\hbar\hat{U}^{\dagger}(t)\frac{d\hat{U}(t)}{dt}\nonumber\\
=&&\hbar
\left(\begin{array}{cccc}
 0                      & \Omega_m/2                &0                        &0\\
\Omega_m/2              & \Delta                    &\Omega_c                 &0\\
0                       & \Omega_c                  & \Delta +\Delta_r        & \Omega_m/2\\
0                       & 0                         & \Omega_m/2              & 2\Delta +\Delta_r \\
\end{array} \right).\nonumber\\
\end{eqnarray}
This transformed Hamiltonian is exactly Eq.(\ref{simple form}). To
understand the relationship between the lifetime of dark periods
and detunings as discussed in Sec. V, we rewrite the Hamiltonian
$\hat{H}'$ in the basis
$\{{|A_e\rangle,|A_g\rangle,|B_e\rangle,|B_g\rangle}\}$. In this
new basis, we find that the Hamiltonian in the resonant case
$\Delta=0$ is actually Eq.(\ref{A-B form}) given by
\begin{eqnarray}
&&\hat{H}_{A-B}=\hat{V}^{\dagger}\hat{H}'\hat{V}\nonumber\\
&&=\hbar
\left(\begin{array}{cccc}
 \Omega_m/2                      & 0                &\Omega_c/2                  &\Omega_c/2 \\
 0                      & -\Omega_m/2               &-\Omega_c/2                 &-\Omega_c/2 \\
\Omega_c/2             &-\Omega_c/2                & \Delta_r+\Omega_m/2        & 0         \\
\Omega_c/2             &-\Omega_c/2                & 0                          & \Delta_r-\Omega_m/2 \\
\end{array} \right),\nonumber\\
\end{eqnarray}
where
\begin{equation}
\hat{V}=\frac{1}{\sqrt{2}}
\left(\begin{array}{cccc}
1                             & 1                                        &0                                     &0 \\
1                             & -1                                       &0                                     &0 \\
0                             &0                                         &1                                     &1 \\
0                             &0                                         & 1                                    &-1\\
\end{array} \right).
\end{equation}\\

For the general off resonance case $\Delta\neq0$, we have
\begin{eqnarray}
&&\hat{H}_{A-B}=\hat{V}^{\dagger}\hat{H}'\hat{V}\nonumber\\
&&=\hbar
\left(\begin{array}{cccc}
 \omega_{Ae}                         & 0                           &\lambda_{Ae,Be}\Omega_c      &\lambda_{Ae,Bg}\Omega_c \\
 0                                   &  \omega_{Ag}                &\lambda_{Ag,Be}\Omega_c      &\lambda_{Ag,Bg}\Omega_c \\
\lambda_{Ae,Be}\Omega_c              &\lambda_{Ag,Be}\Omega_c      &  \omega_{Be}                & 0         \\
\lambda_{Ae,Bg}\Omega_c              &\lambda_{Ag,Bg}\Omega_c      & 0                           & \omega_{Bg}  \\
\end{array} \right),\nonumber\\
\end{eqnarray}
where $\omega_{Ae}=(\Delta+\sqrt{\Omega_m^2+\Delta^2})/2$,
$\omega_{Ag}=(\Delta-\sqrt{\Omega_m^2+\Delta^2})/2$,
$\omega_{Be}=(\Delta+\sqrt{\Omega_m^2+\Delta^2})/2+\Delta+\Delta_r$,
$\omega_{Bg}=(\Delta-\sqrt{\Omega_m^2+\Delta^2})/2+\Delta+\Delta_r$,
$\lambda_{Ae,Be}=\sin\frac{\alpha}{2}\cos\frac{\alpha}{2}$,
$\lambda_{Ae,Bg}=\cos^2\frac{\alpha}{2}$,
$\lambda_{Ag,Be}=-\sin^2\frac{\alpha}{2}$,
$\lambda_{Ag,Bg}=-\sin\frac{\alpha}{2}\cos\frac{\alpha}{2}$, and
\begin{equation}
\hat{V}=
\left(\begin{array}{cccc}
\sin\frac{\alpha}{2}                 & \cos\frac{\alpha}{2}              &0                                &0 \\
\cos\frac{\alpha}{2}                    & -\sin\frac{\alpha}{2}          &0                                &0 \\
0                                    &0                                  &\sin\frac{\alpha}{2}             &\cos\frac{\alpha}{2} \\
0                                    &0                                  &\cos\frac{\alpha}{2}             &-\sin\frac{\alpha}{2}\\
\end{array} \right),
\end{equation}\\
with $\alpha=\arctan(\Omega_m/\Delta)$.


\begin{thebibliography}{0}
\expandafter\ifx\csname natexlab\endcsname\relax\def\natexlab#1{#1}\fi
\expandafter\ifx\csname bibnamefont\endcsname\relax
  \def\bibnamefont#1{#1}\fi
\expandafter\ifx\csname bibfnamefont\endcsname\relax
  \def\bibfnamefont#1{#1}\fi
\expandafter\ifx\csname citenamefont\endcsname\relax
  \def\citenamefont#1{#1}\fi
\expandafter\ifx\csname url\endcsname\relax
  \def\url#1{\texttt{#1}}\fi
\expandafter\ifx\csname urlprefix\endcsname\relax\def\urlprefix{URL }\fi
\providecommand{\bibinfo}[2]{#2}
\providecommand{\eprint}[2][]{\url{#2}}

\end{thebibliography}


\begin{thebibliography}{1}

\bibitem{Mooij} J. E. Mooij, Science \textbf{307},
1210 (2005).

\bibitem{Makhlin} Y. Makhlin, G. Sch\"on, and A. Shnirman, Rev. Mod. Phys \textbf{73},
357 (2001).

\bibitem{Nielsen} M. A. Nielsen and I. L. Chuang, \textit{Quantum Computation and Quantum
Information}(Cambridge Univ. Press, Cambridge, 2000).

\bibitem{Jackel} L. D. Jackel, J. P. Gordon, E. L. Hu, R. E. Howard, L. A. Fetter,
D. M. Tennant, and R. W. Epworth, Phys. Rev. Lett. \textbf{47}, 697 (1981);
R. F. Voss and R. A. Webb, \textit{ibid}., 265 (1981);
R. H. Koch, D. J. Van Harlingen, and J. Clarke, Phys. Rev. Lett. \textbf{47}, 1216 (1981);
J. Clarke, A. N. Cleland, M. H. Devoret, D. Esteve, and J. M. Martinis, Science
 \textbf{239} 992 (1988); J. M. Martinis, M. H. Devoret, and J. Clarke, Phys. Rev. B \textbf{35}, 4682 (1987).

\bibitem{Nakamura} Y. Nakamura, Y. A. Pushkin, and J. S. Tsai, Nature \textbf{398}, 786 (1999);
D. Vion, A. Aassime, A. Cottet, P. Joyez, H. Pothier, C. Urbina, D. Esteve, and M. H. Devoret, Science \textbf{296}, 886 (2002);
I. Chiorescu, Y. Nakamura, C. J. P. M. Harmans, and J. E. Mooij, Science \textbf{299}, 1869 (2003).

\bibitem{Yu} Y. Yu, S. Y. Han, X. Chu, S. I. Chu, and Z. Wang, Science \textbf{296}, 889 (2002).

\bibitem{Martinis} J. M. Martinis, S. Nam, J. Aumentado, and C. Urbina, Phys. Rev. Lett. \textbf{89}, 117901 (2002).

\bibitem{Bohr} N. Bohr, Philos. Mag. \textbf{26}, 476 (1913).

\bibitem{Scully} M. O. Scully and M. S. Zubariry, \textit{Quantum Optics} (Cambridge, 1997).

\bibitem{Orszag} M. Orszag, \textit{Quantum Optics: Including Noise Reduction, Trapped Ions,
Quantum Trajectories, and Decoherence} (Springer-Verlag Berlin Heidelberg, 2000).

\bibitem{Carmichael} H.J. Carmichael, \textit{An Open System Approach to Quantum Optics}, Lecture Notes in Physics (Springer,
Berlin, Heidelberg, 1993).

\bibitem{Plenio} M. B. Plenio and P. L. Knight, Rev. Mod. Phys. \textbf{70}, 101 (1998).

\bibitem{Blatt} R. Blatt and P. Zoller, Eur. J. Phys. \textbf{9}, 250
(1988).

\bibitem{ion} W. Nagourney, J. Sandberg, and H. Dehmelt, Phys. Rev. Lett. \textbf{56}, 2797 (1986);
Th. Sauter, W. Neuhauser, R. Blatt, and P. E. Toschek, Phys. Rev. Lett. \textbf{57}, 1696 (1986);
J. C. Bergquist, R. G. Hulet, W. M. Itano, and D. J. Wineland, Phys. Rev. Lett. \textbf{57}, 1699 (1986).

\bibitem{Helmer} F. Helmer, M. Mariantoni, E. Solano, and F.
Marquardt, Phys. Rev. A \textbf{79}, 052115 (2009).


\bibitem{Romero} G. Romero, J. J. Garcia-Ripoll, and E. Solano, Phys. Rev. Lett. \textbf{102}, 173602
(2009).

\bibitem{Yu2} Y. Yu, S.-L. Zhu, G. Sun, X. Wen, N. Dong, J. Chen, P. Wu, and S. Han, Phys. Rev. Lett.
\textbf{101}, 157001 (2008).
\bibitem{Aumentado} J. Aumentado, M. W. Keller, J. M. Martinis, and M. H. Devoret, Phys. Rev. Lett.
\textbf{92}, 066802 (2004).
\bibitem{Shaw}M. D. Shaw, R. M. Lutchyn, P. Delsing, and P. M. Echternach, Phys. Rev. B
\textbf{78}, 024503 (2008).
\bibitem{Lut} R. M. Lutchyn and L. I. Glazman, Phys. Rev. B
\textbf{75}, 184520 (2007);
R. M. Lutchyn, Phys. Rev. B \textbf{75}, 212501 (2007).

\bibitem{Simmonds} R. W. Simmonds, K. M. Lang, D. A. Hite, S. Nam, D. P. Pappas, and John. M. Martinis,
Phys. Rev. Lett. \textbf{93}, 077003 (2004); K. B. Cooper,
Matthias Steffen, R. McDermott, R. W. Simmonds, Seongshik Oh, D.
A. Hite, D. P. Pappas, and John M. Martinis, Phys. Rev. Lett.
\textbf{93}, 180401 (2004); John M. Martinis, K. B. Cooper, R.
McDermott, Matthias Steffen, Markus Ansmann, K. D. Osborn, K.
Cicak, Seongshik Oh, D. P. Pappas, R. W. Simmonds, and C. C. Yu,
Phys. Rev. Lett. \textbf{95}, 210503 (2005).

\bibitem{Kim}
Z. Kim, V. Zaretskey, Y. Yoon, J. F. Schneiderman, M. D. Shaw, P. M. Echternach, F. C. Wellstood, and B. S. Palmer, Phys. Rev. B
\textbf{78}, 144506 (2008).

\bibitem{Lup}
A. Lupascu, P. Bertet, E.F.C. Driessen, C.J.P.M. Harmans, and J.E.
Mooij, arXiv:0810.0590 (unpublished).

\bibitem{Shnirman}
A. Shnirman, G. Sch\"on, I. Martin, and Y. Makhlin
Phys. Rev. Lett. \textbf{94}, 127002 (2005);
I. Martin, L. Bulaevskii, and A. Shnirman
Phys. Rev. Lett. \textbf{95}, 127002 (2005).

\bibitem{Faoro}
L. Faoro and L. B. Ioffe, Phys. Rev. Lett. \textbf{96}, 047001 (2006).

\bibitem{Clare1}
M. Constantin and C. C. Yu, Phys. Rev. Lett. \textbf{99}, 207001
(2007); M. Constantin, C. C. Yu, and J. M. Martinis, Phys. Rev. B
\textbf{79}, 094520 (2009).

\bibitem{Galperin}
Y. M. Galperin, D. V. Shantsev, J. Bergli, and B. L. Altshuler,
Europhys. Lett. \textbf{71}, 21 (2005).

\bibitem{Clare2}
L.-C. Ku and C. C. Yu, Phys. Rev. B \textbf{72}, 024526 (2005).

\bibitem{Ashhab}
S. Ashhab, J. R. Johansson, and F. Nori, New J. Phys. \textbf{8},
103 (2006).

\bibitem{Neeley} M. Neeley, M. Ansmann, R.C. Bialczak, M. Hofheinz, N. Katz, E. Lucero, A. O. Connell, H.Wang, A. N. Cleland,
and J. M. Martinis, Nature Phys. \textbf{4}, 523 (2008).

\bibitem{Zagoskin} A. M. Zagoskin, S. Ashhab, J. R. Johansson, and F. Nori, Phys. Rev. Lett. \textbf{97}, 077001 (2006).

\bibitem{LinTian} L. Tian and K. Jacobs, Phys. Rev. B \textbf{79}, 144503 (2009).

\bibitem{Zhu} L.-B. Yu, Z.-Y. Xue, Z. D. Wang, Y. Yu, and S.-L. Zhu, arXiv: 0904.1275.

\bibitem{Leggett} A. J. Leggett, in \textit {Chance and Matter}, edited by J. Souletie, J.Vannimenus, and R. Stora (Elsevier, Amsterdam, 1987), p. 395.

\bibitem{Martinis2} J. M. Martinis, M. H. Devoret, and J. Clarke, Phys. Rev. B \textbf {35}, 4682 (1987).

\bibitem{Clarke}     J. Clarke, A. N. Cleland, M. H. Devoret, D. Esteve, and J. M. Martinis,
Science \textbf{239}, 992 (1988).

\bibitem{Martinis3} J. M. Martinis, S. Nam, J. Aumentado, and K. M. Lang, Phys. Rev. B \textbf {67}, 094510 (2003).

\bibitem{Dalibard}  J. Dalibard, Y. Castin, and K. M\"olmer, Phys. Rev. Lett. \textbf{68}, 580 (1992).


\bibitem{Dalibard2} K. M\"olmer, Y. Castin, and J. Dalibard, J. Opt. Soc. Am. A \textbf{10}, 524 (1993).

\bibitem{Carmichael2} L. Tian and H. J. Carmichael, Phys. Rev. A \textbf{46}, 6801 (1992).

\bibitem{Phillips} W. A. Phillips, J. Low Temp. Phys. \textbf{7}, 351 (1972).

\bibitem{Zurek} W. Zurek, Phys. Rev. D \textbf{24}, 1516(1981); \textbf{26}, 1862(1982).

\bibitem{reason1} In the case $\varepsilon>\Delta_0$, the energy basis of TLS is approximated to the position basis.
In this approximation, states $|g\rangle$ and $|e\rangle$
correspond to different ciritical currents. Therefore, at a fixed
bias current, the tunneling rates from $|0g\rangle$($|1g\rangle$)
and $|0e\rangle$($|1e\rangle$) are different. With no loss of
generaility, supposing $|e\rangle$ and $|g\rangle$ correspond to
the lower and higher critical current, respectively, then the
tunneling rate of $|1e\rangle$ is prominent.


\bibitem{Lamb} W. E. Lamb and T. M. Sanders, Phys. Rev. \textbf{119}, 1901(1960);
L. R. Wilcox and W. E. Lamb, Phys. Rev. \textbf{119}, 1915(1960);
J. R. Ackerhalt and B. W. Shore, Phys. Rev. A \textbf{16}, 277(1977).
\end{thebibliography}
\end{document}